\documentclass[12pt,preprint]{aastex}

\begin{document}

\title{Contribution of unresolved point sources to the galactic diffuse emission}

\author{S. Casanova and B. L. Dingus}
\affil{Los Alamos National Laboratory,
    Los Alamos, NM 57545, US}

\begin{abstract}

The detection by the HESS atmospheric Cherenkov telescope of fifteen new 
sources from the Galactic plane makes it possible 
to estimate the contribution of unresolved point sources like those detected by HESS to 
the diffuse Galactic emission measured by EGRET and recently at higher energies by the Milagro 
Collaboration. Assuming that HESS sources have 
all the same intrinsic luminosity, the contribution of this new source 
population can account for most of the Milagro $\gamma$-ray emission at TeV energies and
between 10 and 20 per cent of EGRET diffuse Galactic $\gamma$-ray emission for energies bigger than 10 GeV. Also, by combining the HESS and 
the Milagro results, constraints can be put on the distribution and the luminosities of gamma ray emitters in the Galaxy.

\end{abstract}

\section{Introduction}

The Galactic diffuse $\gamma$-ray emission is believed to be mostly produced in interactions 
of cosmic rays with the matter and the radiation 
fields in the Galaxy, the main production mechanisms being electron non-thermal Bremsstrahlung, 
Inverse Compton scatterings 
off the radiation fields and pion decay processes in inelastic collisions of nuclei and matter. 
Although the standard production mechanisms of $\gamma$-rays \citep{Bertsch:1993} explain 
generally well the spatial and 
energy distribution of the emission below 1 GeV, 
the model does not match EGRET observations of the $\gamma$-ray sky above 1 GeV.\citep{Hunter:1997we} 
The discrepancy between the model and the observations is more pronounced towards the direction of 
the inner Galaxy, where it amounts to 60 per cent of the observed emission. Many possible 
explanations have been proposed to account for the GeV excess. \citet{Aharonian:2000iz} 
have proposed an hardening of the 
cosmic ray proton spectrum. The average proton spectrum in the Galaxy would have to be harder than locally 
measured. \citet{Strong:1999hr,Strong:2004} have suggested 
a renormalization of the electron spectrum in order to fit EGRET data. 
\citet{Kamae:2006} have argued that diffractive processes make the $\gamma$-ray spectrum harder 
than the incident proton spectrum by ~0.05 in power-law index, that the scaling violation produces more $\pi^0$, and, that the inelastic $p-p$ cross section 
has a logarithmic increase with the incident proton energy, so that the combination of these effects could explain about 20 per cent of the "GeV Excess". \citet{deBoer:2005tm} claimed that the features of
the EGRET excess are consistent with the hypothesis of a signal from Dark Matter Annihilation. Recently \citet{Strong:2006} has argued that the excess in the 
diffuse emission measured by EGRET and COMPTEL could be explained by a combination of 
Galactic source populations too dim to be detected by EGRET in addition to the standard production mechanisms. 

In any case a major difficulty when 
studying the diffuse emission is to disantangle the truly diffuse emission 
from that produced by unresolved sources.  

At TeV energies, Milagro has observed diffuse emission from a large region of the Galactic 
plane from Galactic longitude $l$ of ${40}^{o}<l<{100}^{o}$ \citep{Atkins:2005}.   The different theories
to explain the GeV excess make very different predictions about the TeV flux. Recently \citet{Prodanovic} 
have argued that no strong signal of pion decay is seen in the $\gamma$-ray spectrum. The pion decay 
mechanism would not be able to explain the excess above 1 GeV. Therefore, \citet{Prodanovic} 
claimed that the Milagro measurements of the diffuse flux at higher energies revealed a new excess at TeV energies and 
concluded that a different mechanism is required to explain the two excesses, either dark matter decay or 
contribution from unresolved EGRET sources. An extrapolation of the EGRET sources within the range in 
Galactic longitude of the Milagro observation overestimates the diffuse TeV flux measured. 
However, there could be a population of sources undetectable by EGRET, but contributing to both the GeV and TeV excess 
diffuse emission. Recent observations by the TeV observatory, HESS, in fact point to a new class of hard spectrum TeV sources. HESS is an array of 
telescopes using the imaging atmospheric Cherenkov technique to 
detect $\gamma$-rays above 200 GeV. HESS scans the inner part of the Galaxy in the 
region between $-{30}^{o}$ and ${30}^{o}$ degrees in Galactic longitude ($-{30}^{o}<l<{30}^{o}$) and between $-3^{o}$ and $3^o$ degrees in Galactic 
latitude ($-3^o<b<3^o$) with an angular resolution better than 0.1 degrees. Its average sensitivity at the energy threshold 
of 200 GeV is 2 per cent of the Crab flux within $\pm 2^{o}$ degrees in Galactic latitude 
and decreases outside this range. (The flux of the Crab nebula is about $2.4 \times {10}^{-10} \, {\mathrm {cm^{-2}}} \, {\mathrm {s^{-1}}}$ 
for $E>200 \, \mathrm{GeV}$ \citep{Aharonian:2004hegra}). HESS has previously reported the detection of 8 unknown sources close to the Galactic Plane. \citep{Aharonian:2005jn} Recently the total number of 
discovered sources has been updated to 15. \citep{Aharonian:2005kn} Three sources have also 
been detected that are identified with the Galactic Center, 
the SNR RX J1713.7-3946 and G0.9+01. All HESS sources 
are located close to the Galactic Plane, within $\pm 1$ degrees of latitude. \citep{Aharonian:2005kn} The fluxes above 
200 GeV from all sources are fit to power laws, 
$\frac{dN}{dE} = F_{0} \, E^{-\Gamma}$, whose photon indices $\Gamma$ range from 1.8 to 2.8, with a relatively hard mean index 
of 2.32 and RMS of 0.2. 

In the inner Galaxy region observed by HESS the existence of 91 SNRs and 389 pulsars was 
already known at lower energies \citep{Green:2004gr,Manchester}. Many of these 
SNRs and pulsars within the inner Galaxy might emit VHE $\gamma$ rays but only a few $\gamma$ ray sources were previously known.
Although the HESS Collaboration indicates a clear positional coincidence of its sources with a known SNR or pulsar only in a limited 
number of cases, the distribution of Galactic latitude of 
the eighteen VHE $\gamma$ ray sources detected by HESS agrees quite well with the distributions of all SNRs catalogued by \cite{Green:2004gr} and of all pulsars catalogued by \cite{Manchester}. Assuming, for example, that SNRs are a single class of counterparts 
with isotropic luminosity $1.95  \times {10}^{34}  \,  {\mathrm{erg}}  \,  {\mathrm{s^{-1}}}$
to the new HESS sources and taking for them a simple radiative model, \cite{Aharonian:2005kn} found that the location of these sources 
favours a scale height of less than 100 pc, consistent with the hypothesis that these sources are either SNRs or pulsars in 
a massive star forming region. 

The HESS detections of high energy $\gamma$ rays from fifteen new sources has improved significantly the knowledge of both the 
spatial distribution and the spectra and fluxes of VHE $\gamma$-ray galactic sources. These HESS results make it possible to estimate 
with unprecedented precision the contribution of unresolved point sources to the Galactic diffuse emission measured 
up to 100 GeV by EGRET  and recently extended to TeV energies by Milagro.\citep{Atkins:2005} Based on 
HESS results, knowing the sensitivity and the field of view of an experiment, we estimate the number of expected sources and 
their expected VHE $\gamma$-ray flux. Finally we evaluate the contribution of unresolved 
point sources to the diffuse $\gamma$-ray emission for EGRET and Milagro. 

\section{Method}
Under the simplistic assumption that all sources have the same intrinsic energy luminosity $L_{0}$ for $E>E_{th}$, the integral photon 
flux $F$ from a source depends upon the distance $D$ of the detector as
\begin{equation}
F = \frac{\Gamma-2}{\Gamma-1} \, \frac{L_0}{E_{th}} \frac{1}{ 4 \, \pi \, D^2 }  
\label{fluxlum} 
\end{equation}
where $E_{th}$ is the detector threshold energy and $\Gamma$ is the spectral index if the 
$\gamma$-ray emission is a power law.  Inverting Eq.~(\ref{fluxlum}) 
\begin{equation}
D = \sqrt{\frac{k}{F}} \,.
\label{dF}
\end{equation}
where we have defined $k$ as
\begin{equation}
k= \frac{\Gamma-2}{\Gamma-1} \,\frac{L_0}{E_{th}} \frac{1}{ 4 \, \pi} \,.
\end{equation} 
If we assume that all sources are located in the Galactic plane and their surface density, 
which does not have to be constant, is $\sigma_{sources}$, then by differentiating Eq.(\ref{dF}) 
\begin{equation}
dD =  \frac{\sqrt{k}}{2} \,F^{-3/2} \, dF
\end{equation}
we obtain the number of sources per interval of Galactic longitude $dl$ and per interval flux $dF$  
\begin{equation}
d N_{sources} = \sigma_{sources} \,dl  \, D \, dD =  
\frac{1}{2} \, \sigma_{sources} \, \frac{k}{F^2} \, d F \, dl \,.
\label{number}
\end{equation} 
As first case we assume that $\sigma_{sources}$ is constant. Then integrating Eq.(\ref{number}) over all source fluxes above HESS sensitivity of about $ f_{min}=4.8 \times {10}^{-12} \quad {\mathrm {photons}}
\quad {\mathrm {s^{-1}}} \,{\mathrm{{cm}^{-2}}}$ and over the range in longitude $l$ 
surveyed by HESS the number of sources detected by HESS fixes the density of sources in the Galactic plane 
\begin{equation}
N_{sources} = \int_{f_{min}}^{f_{max} } dF \, \int_{l=-\frac{\pi}{6}}^{l=\frac{\pi}{6}} dl \,\,
\frac{k \, \sigma_{sources}}{2 \, F^2} = 18 \,,
\label{intk}
\end{equation}
where the maximum flux detected by HESS was $f_{max}=6.0 \times {10}^{-11} \quad {\mathrm {photons}} \quad 
{\mathrm {s^{-1}}} \,{\mathrm{{cm}^{-2}}}$.

For a given source luminosity $L_0$ from Eq.(\ref{intk})  
\begin{equation}
\sigma_{sources} = \frac{5.7 \times {10}^{-10}}{\pi \, \frac{\Gamma-2}{\Gamma-1} \,\frac{L_0}{E_{th}} \frac{1}{ 4 \, \pi}} \quad {\mathrm {kpc}^{-2} } \,.
\end{equation}
For instance, assuming that the emission follows a power law with spectral index $\Gamma=2.32$, the average spectral index of HESS sources, 
if $L_0 = {10}^{33} \, {\mathrm {erg/s}}$, $\sigma_{sources}= 5.3 \quad {\mathrm {kpc}^{-2}}$, for $L_0 = {10}^{34} \, {\mathrm {erg/s}} $, $\sigma_{sources}=0.5 \quad 
{\mathrm {kpc}^{-2}}$. 
The average distance is 5.6 kpc for this luminosity. This range of surface density is intermediate between the local density of SNRs and pulsars.

The region in the Galaxy where point sources are likely located extends from $D_{min}=0.3$ kpc up to $D_{max}=30$ kpc \citep{Swordy}. 
Assuming that the flux from each source depends only upon the distance from the detector $F=F(D)$ and not on the type of source, 
HESS total flux per steradian from resolved and unresolved sources is  
\begin{eqnarray*}
F(E>200 \, GeV, \, |l|<30) &=& \int_{D_{min}}^{D_{max}} F(D) \,  dN_{sources} \\[2mm]
&=&   \int_{D_{min}}^{D_{max}}  \,\int_{l=-\frac{\pi}{6}}^{l=\frac{\pi}{6}} \, \frac{k \, \sigma_{sources}}{D^2} \, dl \, D \, 
dD \\[2mm]
&=& 8.0 \times {10}^{-9} \quad {\mathrm {photons}} \quad {\mathrm{{cm}^{-2}}} \, {\mathrm {s^{-1}}} \,
{\mathrm{{sr}^{-1}}} \,.
\end{eqnarray*}
\begin{equation}
\label{fluxtotal}
\end{equation}
In Eq.~\ref{fluxtotal} the total integral flux is already divided by the HESS field of view. This flux is independent of $L_{0}$.  

The surface density of $\gamma$-ray sources 
is not likely to be constant. The distribution is probably similar to that of 
$\gamma$-ray candidate sources such as pulsars or supernova remnants which are
more prominent in the inner galaxy. However, 
the ratio of $\gamma$ emitting pulsars versus SNRs is unknown. 
The two extreme cases of all HESS sources being either pulsars or SNRs 
are considered. $\sigma_{sources}$ in Eq.(\ref{number}) will be either 
the surface density of SNRs or the surface density of pulsars in the Galactic Plane, 
as observed in the radio wavelengths. We do not consider 
the latitude distributions of pulsars and SNRs because most point sources are 
located close to the plane of the Galaxy, within a region ($-2^{o}<b<2^{o}$) which extends 
from $D_{min}=0.3$ kpc up to $D_{max}=30$ kpc \citep{Swordy}. In terms of the 
heliocentric distance $D$ and the longitude angle $l$ the galactocentric distance, $r$, is
\begin{equation}
r = \sqrt{r_0^2 + D^2 - 2 \, r_0 \, D \, cos(l)}
\end{equation}
with $r_0=8.5$ ${\mathrm{kpc}}$, which is the distance from the Earth to the Galactic center. 
The pulsar surface density $\sigma_{pulsar}(r)$, plotted in Fig.1, is fitted by 
the following shifted Gamma function \citep{Yusifov:2004fr,Lorimer:2004}
\begin{equation}
\sigma_{pulsar}(r) = {\sigma_0}_{p} \, {(\frac{x}{x_0})}^a \, e^{[-b \, (\frac{x-x_0}{x_0})]}
\end{equation}
where $x=r+r_1$ and $x_0 = r_0+r_1$, ${\sigma_0}_p = 37.6\pm 1.9$ ${\mathrm {kpc^{-2}}}$, 
$a=1.64\pm 0.11$, $b=4.01\pm0.24$ and $r_1=0.55\pm0.10$ ${\mathrm{kpc}}$. The SNR surface density, 
plotted in Fig.1, is \citep{Green:2004gr,Case:1998qg}
\begin{eqnarray}
\sigma_{SNR}(r) = \left\{ \begin{array}{c@{\hspace{12mm}}l}
{\sigma_0}_{snr}  \,sin(\frac{\pi \,r}{r_2}+\theta_0) \,  e^{-\beta r}  &  \quad {\mathrm for} \quad {\mathrm r} < 16.8 \\[2mm]
0 &  \quad {\mathrm for} \quad {\mathrm r} > 16.8
\end{array} \right. 
\label{sigmasnr}
\end{eqnarray}
with ${\sigma_0}_{snr} = 1.96\pm 1.38$ ${\mathrm {kpc^{-2}}}$, $r_2=17.2\pm 1.9$ ${\mathrm {kpc}}$, 
$\theta_0=0.08\pm0.33$ and $\beta=0.13\pm0.08$. Integrating Eq.(\ref{number}) over all pulsar fluxes above HESS sensitivity, $f_{min}$, and below $f_{max}$, 
the greatest flux from a source detected by HESS, and assuming a given source luminosity $L_0$, the number 
of sources resolved by HESS in the region ($-{30}^{o}<l<{30}^{o}$) fixes the maximum fraction of 
pulsars that are $\gamma$-ray emitters
\begin{eqnarray*}
N_{HESS} &=& \int   dD \, D  \, \int  d l \, \sigma_{pulsar}(l,D) = 18 \\[2mm]
&=&  \frac{1}{2} \,\int_{f_{min}}^{f_{max}} dF \,  \frac{k}{F^2}  \, 
\int_{l=-\frac{\pi}{6}}^{l=\frac{\pi}{6}} d l \, \sigma_{pulsar}(l,F) = 18  \,.
\end{eqnarray*}
\begin{equation}
\label{sensitivity}
\end{equation}
\begin{figure}[ht]
\begin{center} 
\includegraphics[angle=0,width=8cm]{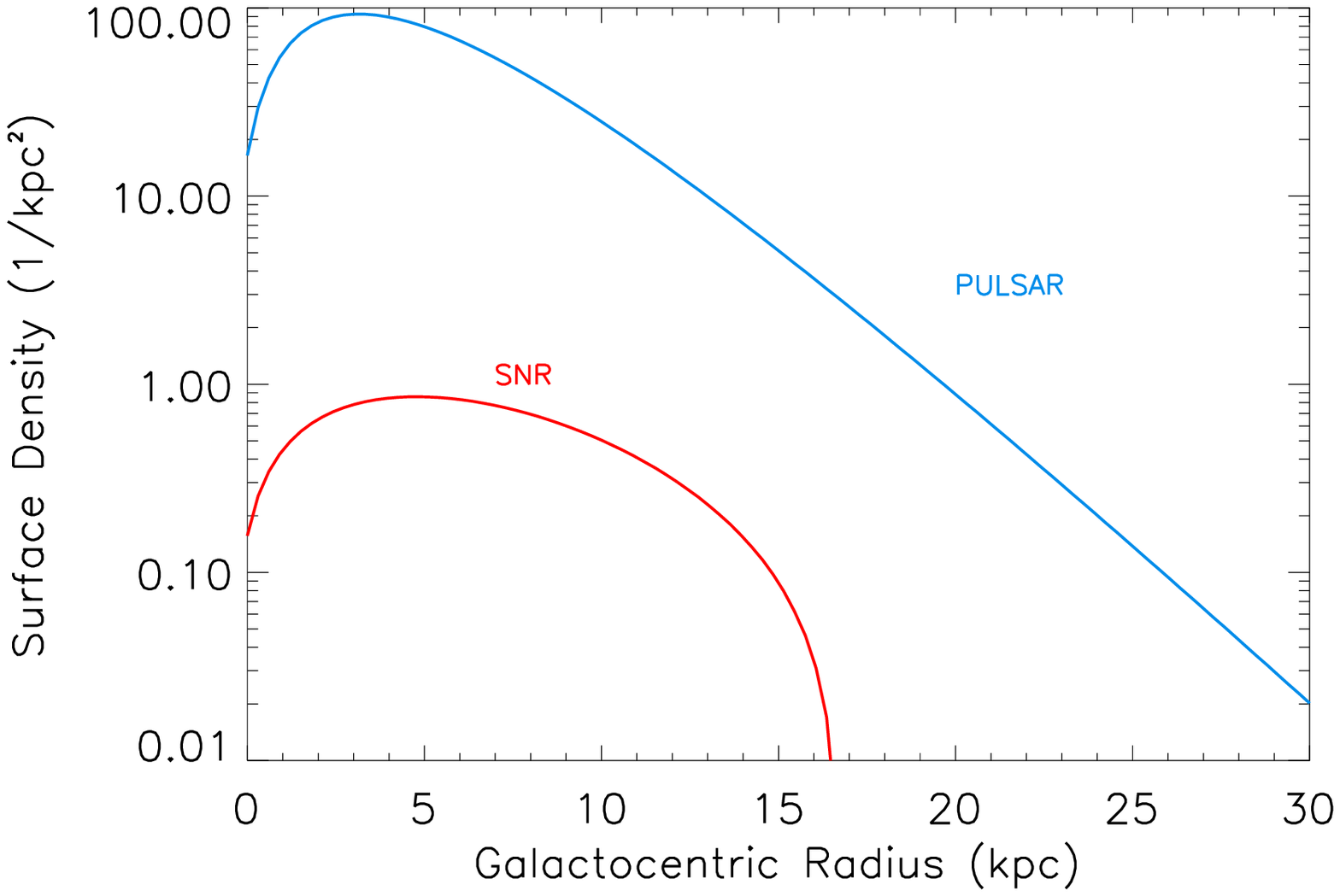}
\caption{{ Pulsar and SNR surface distributions versus galactocentric radius $r$ 
from \citet{Yusifov:2004fr} and \citet{Lorimer:2004} and from \citet{Green:2004gr} 
and \citet{Case:1998qg}, respectively.} \label{fig1}}
\end{center}
\end{figure}
If HESS population of sources consists entirely of pulsars, the total flux 
from resolved and unresolved sources divided over HESS field of view is 
\begin{eqnarray*}
F_{pulsar}(E>200 \, GeV, \, |l|<30) &=& \int \, F(D) \, dN_{pulsar} \\[2mm] 
&=& \int_{D_{min}}^{D_{max}} dD \, D \,  \frac{k}{D^2}  \, \int_{l=-\frac{\pi}{6}}^{l=\frac{\pi}{6}} d l \, \sigma_{pulsar}(l,D)  \\[2mm]
&=&  1.0 \times {10}^{-8} \quad {\mathrm {photons}}  \quad {\mathrm{cm^{-2}}} \quad {\mathrm {s^{-1}}} \quad 
{\mathrm {{sr}^{-1}}} \,. 
\end{eqnarray*}
\begin{equation}
\label{eqn:prima}
\end{equation}
If the all radio loud pulsars emit $\gamma$-rays then the luminosity predicted in Eq.~(\ref{sensitivity}) is $1.2 \times {10}^{32}$ ergs/s and the average distance would be 0.5 kpc. Though the ratio between $\gamma$-ray loud 
versus $\gamma$-ray quiet pulsars is uncertain. If we suppose that only five per cent of the radio loud pulsar emit $\gamma$-rays, then the 
\begin{equation}
F_{pulsar}(E>200 \, GeV, \, |l|<30) = 7.0 \times {10}^{-9} \quad {\mathrm {photons}}  \quad {\mathrm{cm^{-2}}} \quad {\mathrm {s^{-1}}} \quad 
{\mathrm {{sr}^{-1}}} \, ,
\label{eqn:seconda}
\end{equation}
the source luminosity predicted would be $1.7 \times {10}^{33}$ ergs/s and the average distance for HESS sources 1.8 kpc. If 1 per cent of radio loud pulsars emit TeV $\gamma$s then
\begin{equation}
F_{pulsar}(E>200 \, GeV, \, |l|<30) =  5.4 \times {10}^{-9}  \quad {\mathrm {photons}}  \quad {\mathrm{cm^{-2}}} 
\quad {\mathrm {s^{-1}}} \quad {\mathrm {{sr}^{-1}}} \, .
\end{equation}
the source luminosity predicted would be $6.5 \times {10}^{33}$ ergs/s and the 
average distance for HESS sources 3.5 kpc. 

\begin{figure}[ht]
\begin{center}
\includegraphics[angle=0,width=8cm]{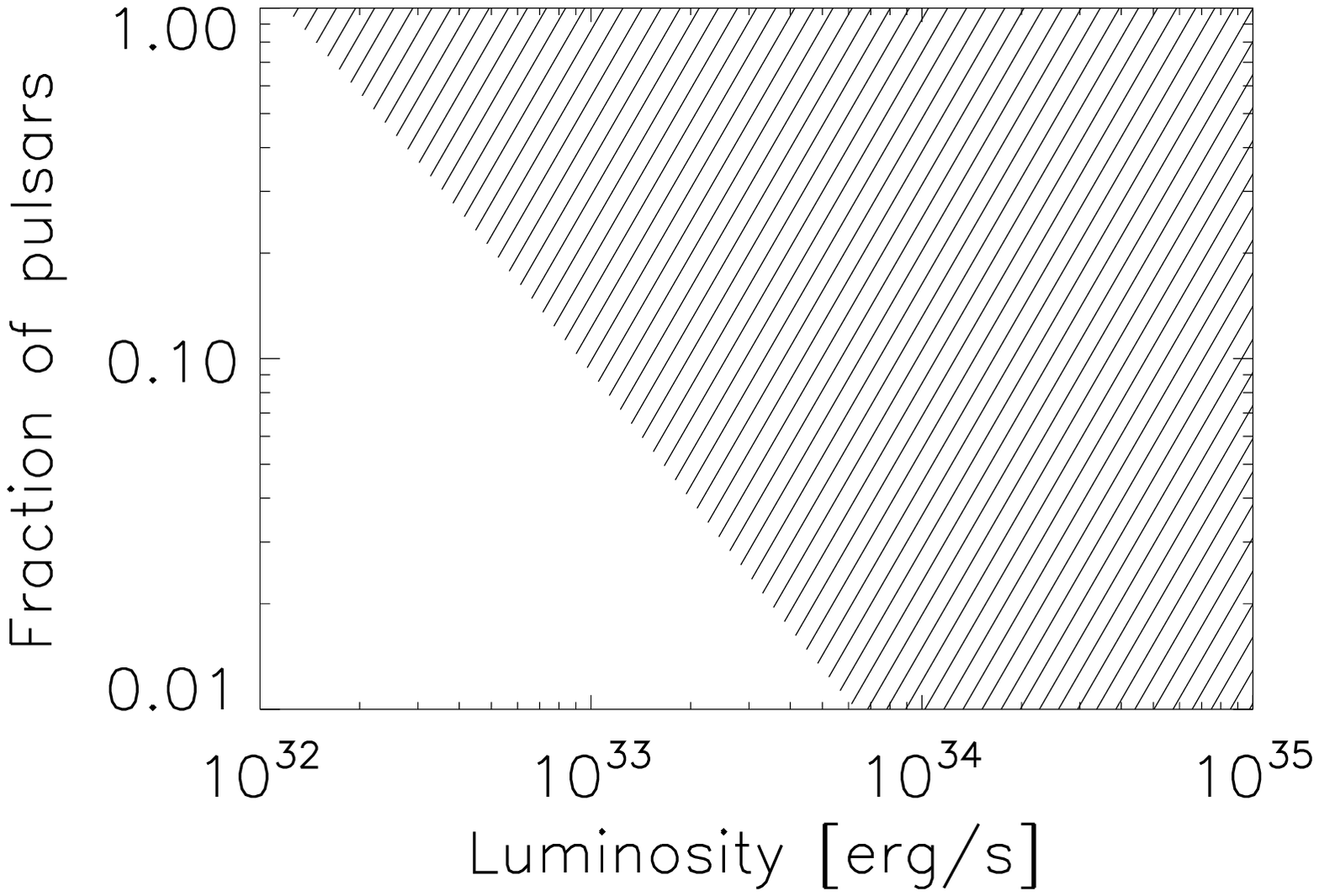}
\includegraphics[angle=0,width=8cm]{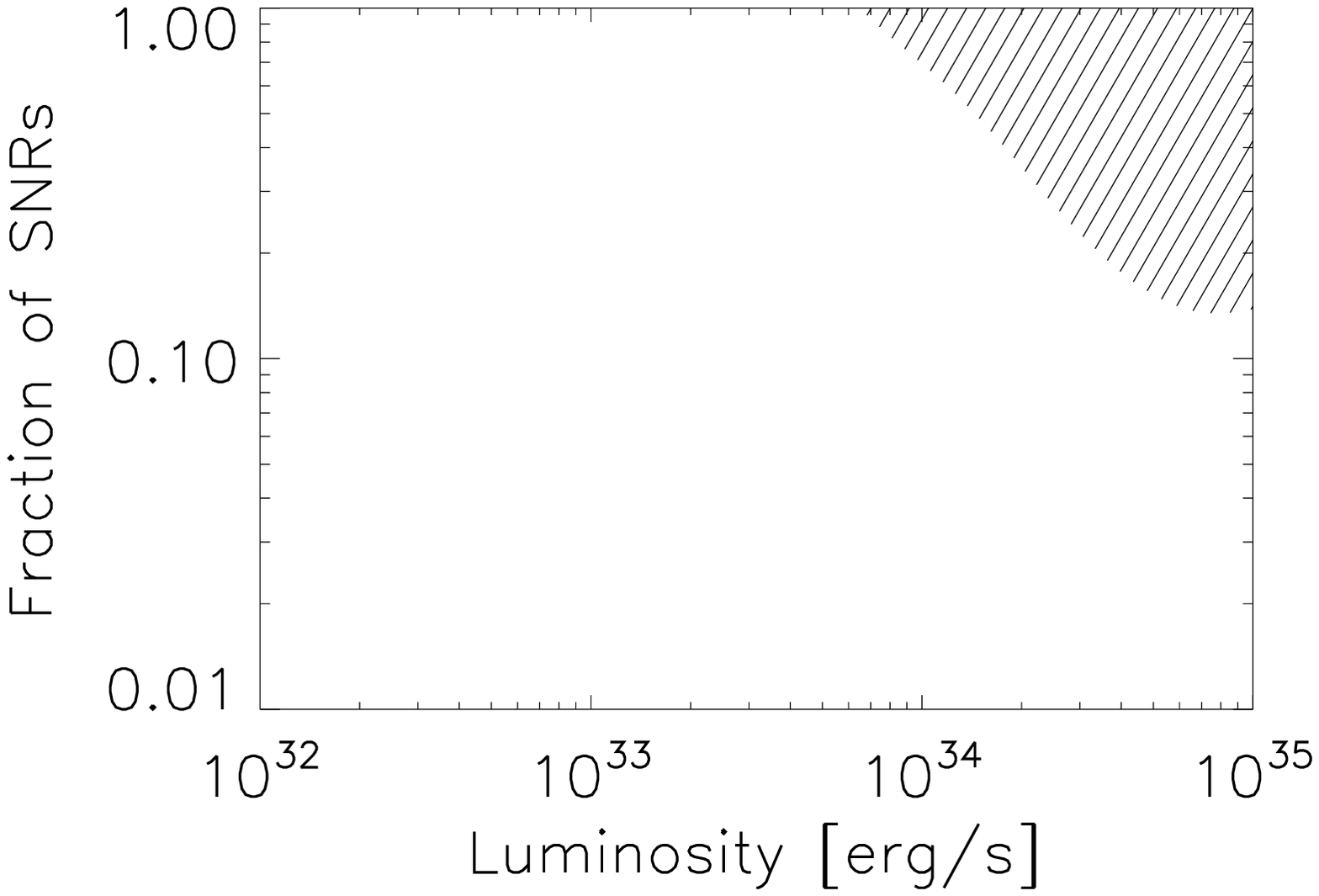}
\caption{{Values of luminosity $L_{0}$ and fraction of $\gamma$-ray emitting pulsars or SNRs allowed by the condition that HESS detects 18 sources. The shaded 
region is not allowed. \label{fig2} }}
\end{center}
\end{figure}

If instead all HESS sources belong to the SNR population the total flux would be 
\begin{eqnarray*}
{F_{snr}}(E>200 \, GeV, \, |l|< 30) &=& \int \, F(D) \, dN_{SNR} \\[2mm] 
&=&\int_{D_{min}}^{D_{max}} dD \, D \,  \frac{k}{D^2}  \, \int_{l=-\frac{\pi}{6}}^{l=\frac{\pi}{6}} d l \, \sigma_{SNR}(l,D)  \\[2mm]
&=& 6.6 \times {10}^{-9} \quad {\mathrm {photons}}  \quad {\mathrm{cm^{-2}}} 
\quad {\mathrm {s^{-1}}} \quad {\mathrm {{sr}^{-1}}} \, ,
\end{eqnarray*}
\begin{equation}
\label{eqn:terza}
\end{equation}
the source luminosity predicted for HESS sources would be $6.5 \times {10}^{33}$ ergs/s and their average distance 3.5 kpc. Fig~\ref{fig2} shows the values of luminosity $L_{0}$ and fraction of $\gamma$-ray emitting pulsars or SNRs, allowed by the condition that HESS detects 18 sources.  

In the HESS catalogue of fifteen new sources 
8 sources have SNRs as possible counterparts, and 4 of these new sources are potentially associated to pulsar wind nebulae.
So pulsars and SNRS both occur but their surface densities are likely not so different.
\begin{figure}[ht]
\begin{center}
\includegraphics[angle=0,width=8cm]{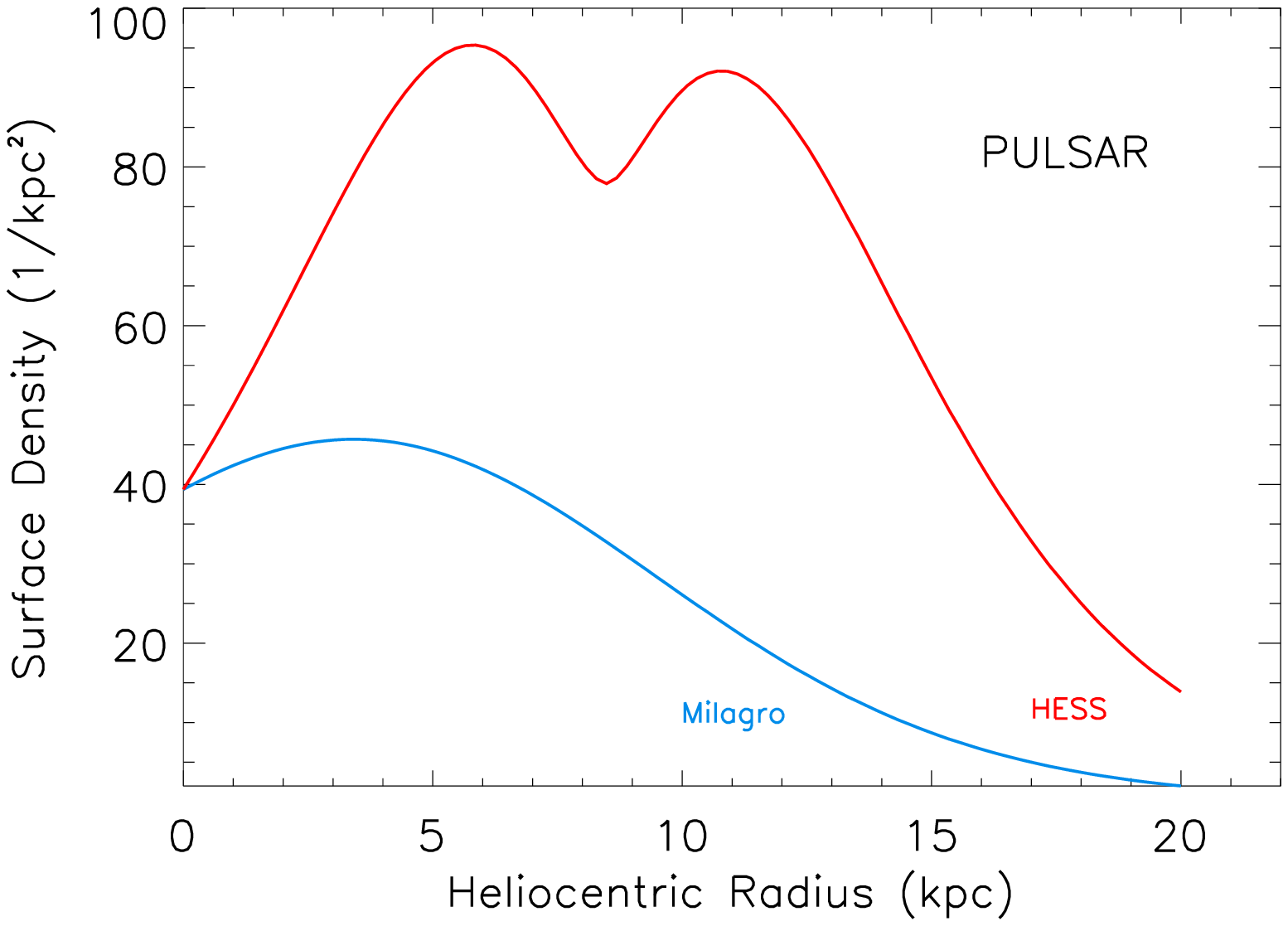}
\includegraphics[angle=0,width=8cm]{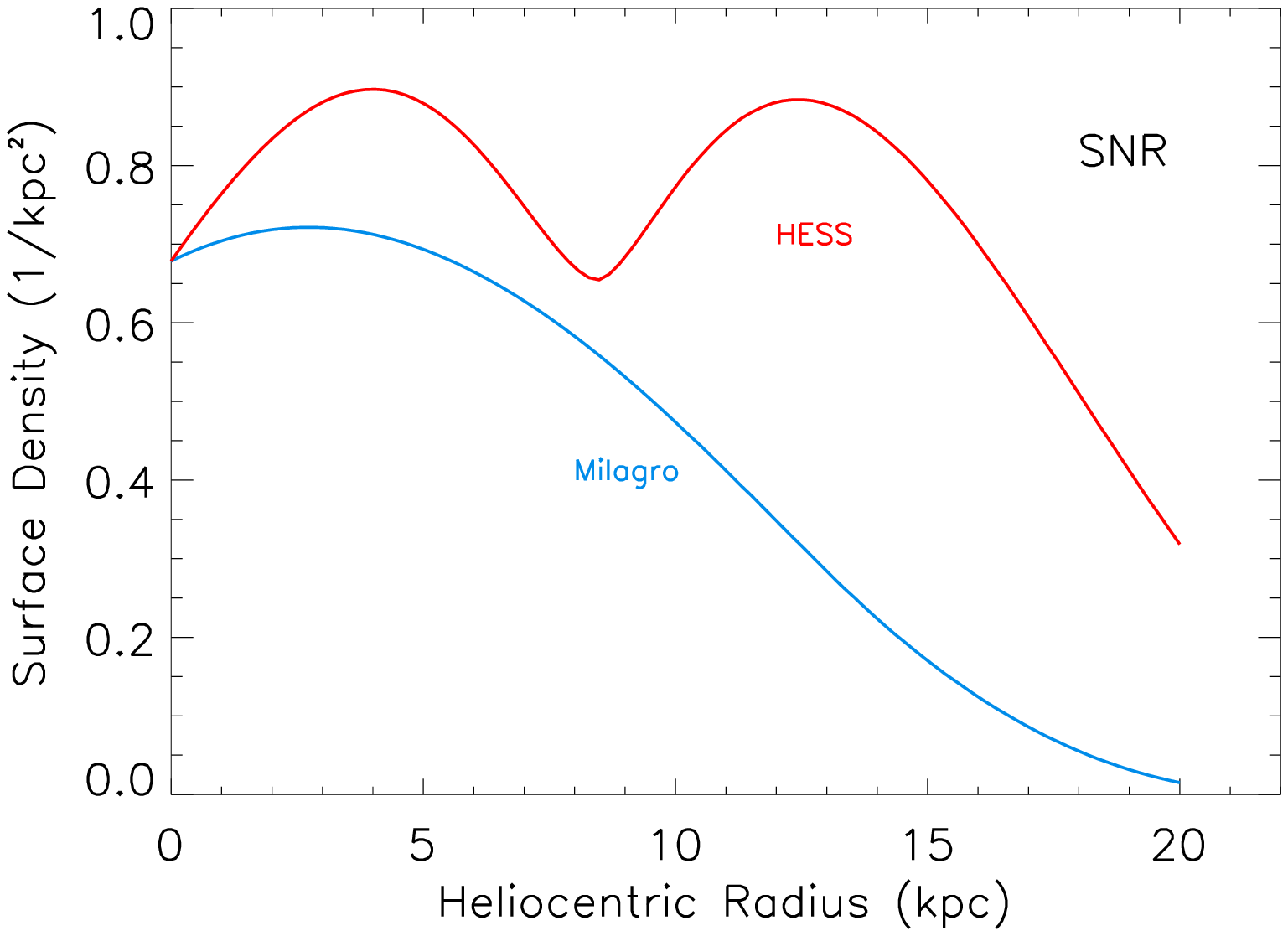}
\caption{Pulsar and SNR surface distributions versus the heliocentric distance D for the regions 
of the Galactic plane surveyed 
by Milagro (~${40}^{o}<l<{100}^{o}$~) and by HESS (~$-{30}^{o}<l<{30}^{o}$~).\label{fig3}}
\end{center}
\end{figure} 

\section{Contribution of unresolved sources to Milagro diffuse galactic emission}
Milagro is a water Cherenkov telescope, which has surveyed most of the northern sky at TeV energies 
\citep{Atkins:2000,Atkins:2003}. Milagro has measured TeV energy fluxes from the Crab 
\citep{Atkins:2004jf} and the Cygnus Region \citep{Smith} and, recently the diffuse 
Galactic $\gamma$-ray emission up to TeV energies. 
Milagro Galactic diffuse emission is $(7.3  \pm 1.5  \pm 2.3)  \times {10}^{-11} \quad {\mathrm {{photons}}}
\quad {\mathrm {s^{-1}}} \quad {\mathrm{{cm}^{-2}}} \quad {\mathrm{{sr}^{-1}}}$ for $E>3.5$ TeV 
in the region ${40}^{o}<l<{100}^{o}$. \citep{Atkins:2005} The detection of TeV $\gamma$-ray diffuse emission 
is of fundamental importance in solving the 
long standing problem of the production of high energy $\gamma$ radiation in our Galaxy. In fact the 
Milagro result seems to exclude an additional hard spectrum component continuing to above 10 TeV to 
explain the 60 per cent excess of EGRET flux compared to $\pi_{0}$-decay 
production mechanism due to the local cosmic ray flux. \citep{Hunter:1997we} In order to discriminate 
among different production mechanisms of the diffuse radiation, it is important to estimate the 
amount of the measured diffuse emission which could arise from unresolved pointlike sources. The Milagro 
sensitivity to point sources is about ${10}^{-11}  \quad {\mathrm {{photons}}}\quad {\mathrm {s^{-1}}} 
\quad {\mathrm{{cm}^{-2}}}  \quad {\mathrm for} \quad E > 1 TeV$  and, in fact, it is of the order of the 
brightest integral source flux detected by HESS. However, part of the diffuse emission measured by Milagro 
is emission from unresolved HESS-like pointlike sources.

The integral flux expected from the Milagro field of view from HESS-like sources assuming a pulsar 
shaped surface density and source luminosity $L_{0}=1.7 \times {10}^{33} \quad {\mathrm{erg/s}}$ is
\begin{eqnarray*}
F_{pulsar}(E>200 \, GeV, \, 40<l<100) &=& \int \, F(D) \, dN_{pulsar}  \\[2mm]
&=& \int_{0.3 \, kpc}^{30 \, kpc} dD \, D \,  \frac{k}{D^2} \, \int_{l=\frac{2\pi}{9}}^{l=\frac{5\pi}{9}} d l \, \sigma_{pulsars}(l,D) \\[2mm] 
&=&  2.4 \times {10}^{-9} \quad {\mathrm{{photons}}}
\quad {\mathrm {s^{-1}}} \,{\mathrm{{cm}^{-2}}} {\mathrm {sr}^{-1}} \,, 
\end{eqnarray*}
\begin{equation}
\label{eqn:terzabis}
\end{equation}
while from HESS-like sources assuming a SNR shaped surface density and $L_{0}=6.5 \times {10}^{33} 
\quad {\mathrm{erg/s}} $ is
\begin{eqnarray*}
F_{snr}(E>200 \, GeV, \, 40<l<100)  &=& \int \, F(D) \, dN_{SNR}\\[2mm] 
&=& \int_{0.3 \, kpc}^{30 \, kpc} dD \, D \,  \frac{k}{D^2}  \, 
\int_{l=\frac{2\pi}{9}}^{l=\frac{5\pi}{9}} d l \, \sigma_{SNR}(l,D) \\[2mm] 
&=& 3.1 \times {10}^{-9}   \quad {\mathrm {{photons}}}
\quad {\mathrm {s^{-1}}} \,{\mathrm{{cm}^{-2}}} {\mathrm {sr}^{-1}} \,, 
\end{eqnarray*}
\begin{equation}
\label{eqn:quarta}
\end{equation}
The fluxes in Eq.~(\ref{eqn:terzabis}) and (\ref{eqn:quarta}) are lower than HESS flux 
from resolved and unresolved sources. In fact as shown in Fig.3 HESS observes the inner 
region of the Galaxy, whereas Milagro field of view is more spread toward the outer regions 
in the Galaxy, where the number of sources is substantially lower, so a lower contribution 
from unresolved sources is expected for Milagro's region of the Galactic plane than for HESS. 

\begin{figure}[h]
\begin{center}
\includegraphics[angle=0,width=7.5cm]{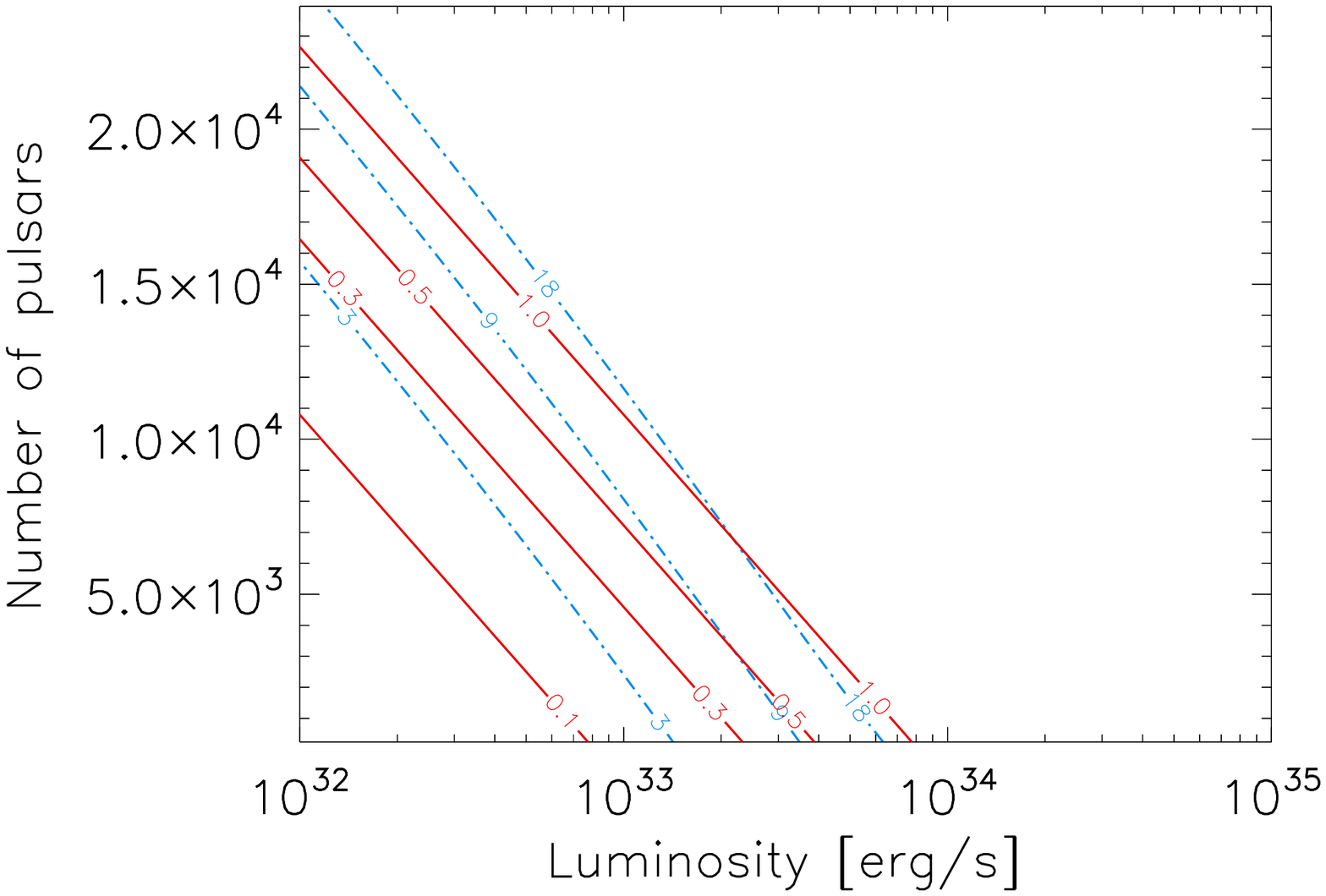}
\includegraphics[angle=0,width=7.5cm]{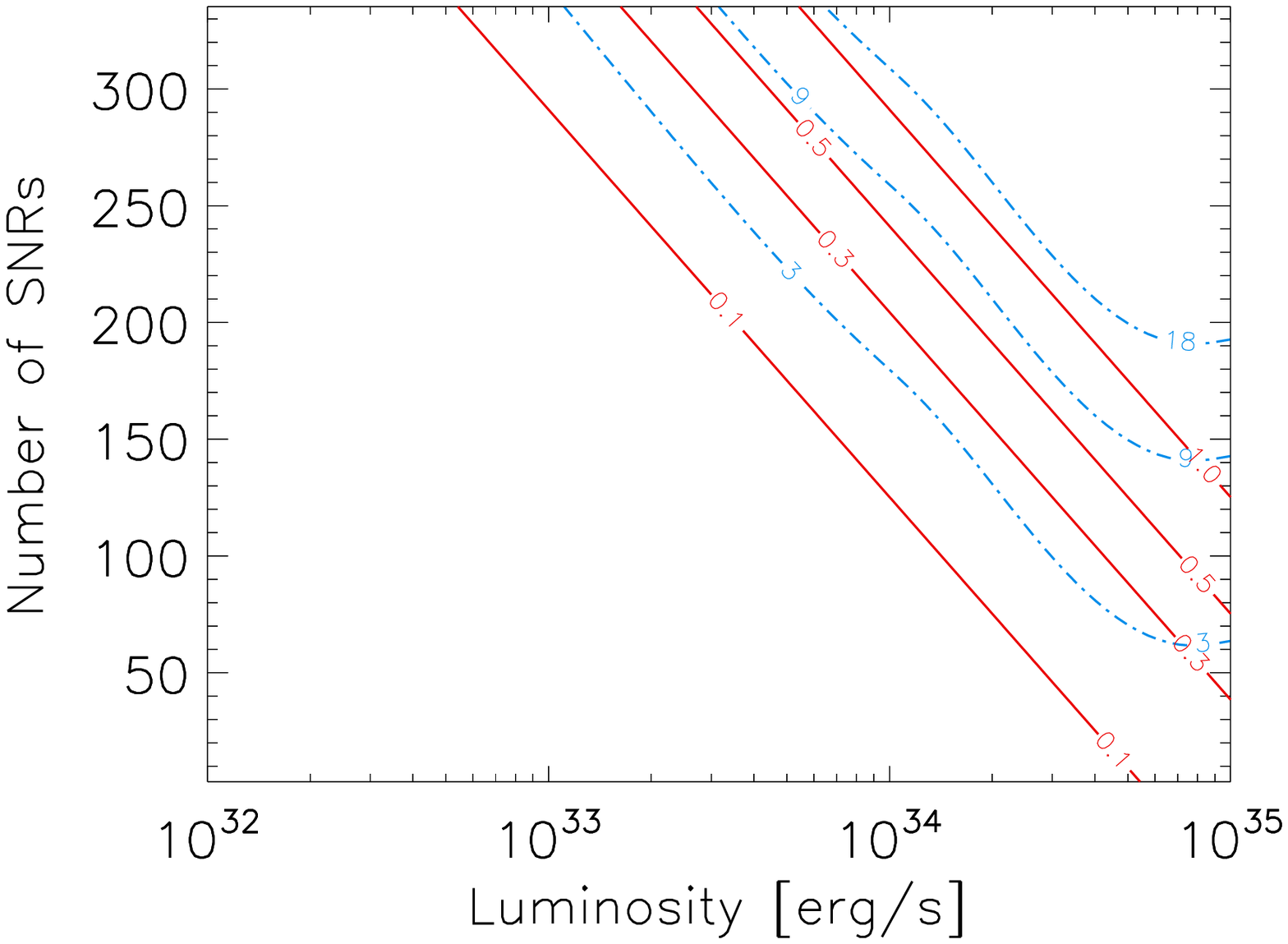}
\caption{The red lines represent the fraction of the measured Milagro flux which resolved and unresolved sources contribute to for different $L_{0}$ and 
for different total number of Galactic SNRs and pulsars which are $\gamma$-ray loud. The lines in blue represent the conditions for HESS 
to detect 3, 9 and 18 sources, respectively. \label{fig4}  }
\end{center}
\end{figure}

Milagro Galactic diffuse emission is measured for a threshold energy of about 3.5 TeV, 
whereas the fluxes from unresolved sources calculated in Eq.(\ref{eqn:terzabis}) and (\ref{eqn:quarta})
refer to HESS threshold energy of 200 GeV. In order to estimate the contribution of unresolved 
sources to Milagro diffuse emission we correct the flux in Eq.(\ref{eqn:terzabis}) and 
(\ref{eqn:quarta}) for the different threshold energy. All HESS sources are 
fitted by a power law spectrum 
\begin{equation}
\Phi(E,\Gamma) = \Phi_0 \, {( \frac{E}{1 TeV} )}^{-{\Gamma}} \,,
\end{equation}
and the average spectral index $\Gamma=2.32$. 
The integral flux from unresolved pulsars if 5 per cent of radio loud pulsars emit $\gamma$-rays corrected for Milagro threshold becomes 
\begin{eqnarray*}
F_{pulsar}(E>3.5 \, TeV, \, 40<l<100) &=&  F_{pulsar}(E>200 \, GeV, \, 40<l<100) \\[2mm]
& \times &{(\frac{E_{Milagro}}{E_{HESS}})}^{(-\Gamma+1)} \\[2mm]
&& =  5.5 \times {10}^{-11}  \quad {\mathrm{{photons}}}
\quad {\mathrm {s^{-1}}} \,{\mathrm{{cm}^{-2}}} {\mathrm {sr}^{-1}}  
\end{eqnarray*}
\begin{equation}
\label{fluxtotaltris}
\end{equation} 
If all HESS sources coincide with SNRs 
\begin{eqnarray*}
F_{snr}(E>3.5 \, TeV, \, 40<l<100) &=&  F_{pulsar}(E>200 \, GeV, \, 40<l<100) \\[2mm]
& \times &{(\frac{E_{Milagro}}{E_{HESS}})}^{(-\Gamma+1)} \\[2mm]
&=&7.1 \times {10}^{-11}  \quad {\mathrm {{photons}}}
\quad {\mathrm {s^{-1}}} \,{\mathrm{{cm}^{-2}}} {\mathrm {sr}^{-1}}  
\end{eqnarray*}
\begin{equation}
\label{fluxtotaltris2}
\end{equation} 
where $E_{Milagro}$ and $E_{hess}$ are Milagro and HESS energy threshold, respectively. The contribution from unresolved 
HESS-like sources to Milagro diffuse emission is comparable to the diffuse emission itself. 
Despite the uncertainty of this calculation, concerning especially the ratio bewteen $\gamma$-ray loud pulsars and SNRs, the result seems to indicate 
that unresolved point sources contribute significantly to Milagro diffuse $\gamma$-ray emission. Fig.~\ref{fig4} shows the fraction of the measured Milagro flux 
which unresolved sources contribute to for different $L_{0}$ and for various total numbers of Galactic SNRs and pulsars. The lines in blue represent 
the conditions for HESS to detect 3, 9 and 18 sources, respectively.  

\section{Contribution of HESS sources to EGRET diffuse galactic emission}

EGRET surveyed the $\gamma$-ray sky in the energy region between 30 MeV and 100 GeV. 
The Third EGRET Catalog lists 80 sources for latitudes less than $\pm 10$
degrees, of which 5 are pulsars, one is a solar flare and 74 are unidentified sources. Studies of the spectral characteristics
\citep{Merck} and of the luminosity function \citep{Mukherjee:1995zt} of EGRET unidentified sources show as unlikely that even a small
part of these sources are pulsars or SNRs. The contribution of these sources to EGRET flux was evaluated by EGRET Collaboration and
should be less than 10 per cent. Recently \citet{Strong:2006} showed that 
the contribution of unresolved sources with luminosity $L (E>100 {\rm MeV}) < 1 \times 
{10}^{35} \quad {\mathrm {photons}} \quad {\mathrm {s^{-1}}}$, which are 
too 
dim to be detected by EGRET, could provide up to 5, 10 per cent of EGRET emission.  Also he showed that combinations of different 
source populations could explain the excess $\gamma$-ray emission measured 
by COMPTEL and EGRET at MeV and GeV energies, respectively. 

The HESS new population of sources does not seem to coincide with
EGRET Catalogue \citep{Aharonian:2005kn}. From \citet{Gonthier:2001wf} the EGRET point source sensitivity in the Galactic plane 
is about $1.6 \times {10}^{-7}  \quad {\mathrm {photons}} \quad {\mathrm {s^{-1}}} \quad {\mathrm{{cm}^{-2}}} \quad {\mathrm {for}} 
\quad {\mathrm{E}}>100 \, {\mathrm{MeV}}$. If we can believe that HESS TeV source fluxes continue down 
to GeV energies, ten of these sources would have been detectable by EGRET if extrapolated 
without any cut-off to 100 MeV. The fact that only a few of HESS new sources 
are spatially consistent with the 95 per cent positional error of unidentified EGRET sources ( HESS J1640-465, HESS J1745-303 and HESS J1825-137 
for which a simple power law estrapolation of the spectrum to 1 GeV gives EGRET flux ) suggests that the simple estrapolation of HESS 
source fluxes to EGRET energies is probably not correct. These sources might have a cut-off close to 10 GeV.  All these sources will be detectable by GLAST, which will have a flux threshold of about  
$5.0 \times {10}^{-9} \quad {\mathrm {photons}} \quad {\mathrm {s^{-1}}} \quad {\mathrm{{cm}^{-2}}}$.   
The sum of the integral fluxes of the three HESS sources, which could coincide with EGRET unidentified sources, 
is  $ 1.1 \times {10}^{-6} \quad {{\mathrm{photons}}}
\quad {\mathrm {s^{-1}}} \quad {\mathrm{{cm}^{-2}}} \quad {\mathrm {sr}^{-1}}$ at 1.4 GeV, about one per cent 
of the EGRET diffuse flux. 

Following the method introduced in Chapter 2 we evaluate the contribution due to HESS-like sources 
to EGRET diffuse emission $F_{sources}(E>E_{0}, \, -{30}^{o}<l<{30}^{o})$ at different energies $E_{0}$ above 
0.50 GeV and for the two cases in which either all SNRs or 5 per cent of the radio loud pulsars 
contribute. In Table~1 the first column gives the energy $E_0$ above which the integral flux $F_{sources}(E>E_{0}, \, -{30}^{o}<l<{30}^{o})$ is evaluated. 
The measured EGRET integral flux for $E>E_0$ is given in the second column. The contribution of HESS-like pulsars or SNRs is given in columns third and 
fourth, respectively. As shown in Table~1 the contribution from HESS-like sources can account 
for about 10 per cent of the GeV excess. This contribution becomes more important at higher 
energies. At about 85 GeV the contribution from either HESS-like pulsars or SNRs is about 
20 per cent of the diffuse flux.

\begin{center}
\begin{figure}[ht]
\begin{tabular}{|c|c|c|c|}
\hline
ENERGY (GeV)    &  EGRET FLUX  &     5 \% PRS FLUX \, ( \% EGRET)    &  SNR FLUX \, (\% EGRET) \\[2mm] \hline
$0.75$     &  $7.9 \times {10}^{-5}$    &   $ 6.7 \times {10}^{-6} \quad 
(9\%)$   &    $6.3 \times {10}^{-6}  \quad (8 \%)$  \\[2mm]
\hline
$1.5$      &  $3.6 \times {10}^{-5}$    &    $ 2.7 \times {10}^{-6} \quad 
(8 \%)$   &    $2.5  \times {10}^{-6} \quad ( 7 \%)$    \\[2mm]
\hline
$3.0$      &  $1.3 \times {10}^{-5}$    &    $ 1.1 \times {10}^{-6} \quad 
( 8\%)$ &    $1.0  \times {10}^{-6} \quad ( 8 \%)$ \\[2mm]
\hline
$7.0$      &  $2.3 \times {10}^{-6}$     &       $3.5  \times {10}^{-7} 
\quad (15 \%)$     &    $3.3 \times {10}^{-7}  \quad ( 14 \%)$ \\[2mm]
\hline
$15.0$     &  $1.4 \times {10}^{-6}$     &      $1.3 \times {10}^{-7}  \quad (9 \%)$    &
$1.2 \times {10}^{-7}  \quad (9 \%)$ \\[2mm]
\hline
$35.0$     &  $4.4 \times {10}^{-7}$     &      $4.2 \times {10}^{-8}  \quad (10 \%)$     &    
$4.0 \times {10}^{-8}  \quad (9 \%)$  \\[2mm]
\hline  
$85.0$     &  $6.4\times {10}^{-8}$     &      $1.3\times {10}^{-8}  \quad ( 20 \%)$    &    
$ 1.2\times {10}^{-8}  \quad (19 \%)$ \\[2mm]
\hline
\end{tabular}
\caption{Table 1. {Integral fluxes for EGRET measured diffuse $\gamma$-ray spectrum of the inner Galaxy region, ($-30^{o}<l<30^{o}$ and 
$-5^{o}<b<5^{o}$) compared with the contribution from HESS-like pulsars and SNRs for different energies. All fluxes are in units of in $\quad {\mathrm {{photons}}}
\quad {\mathrm {s^{-1}}} \quad {\mathrm{{cm}^{-2}}} \quad {\mathrm {sr}^{-1}}$. EGRET fluxes are from \citet{Strong:2005}  
\label{Table1}.}}
\end{figure}
\end{center}

\section{Conclusions}
Using HESS new sample of Galactic $\gamma$-ray emitters we estimated the contribution of unresolved sources 
to the diffuse emission measured by EGRET and by Milagro. We find that the contribution of unresolved HESS-like sources to the
diffuse $\gamma$ ray emission is large at energies above 100 GeV, whereas it amounts to about 10 per cent of 
EGRET measurements of the diffuse flux at GeV energies. Since pulsars and SNRs are distributed mostly 
in the Galactic plane, their unresolved fluxes will mainly contribute to the low latitude diffuse emission. Thus we expect that the total flux due to unresolved sources 
contributes less at higher latitudes. 

The main uncertainties of this calculation consist in assuming that all sources are monoluminous 
and the distribution of $\gamma$-ray sources follows the distribution of either 
pulsars or SNRs. Of the 1509 pulsars catalogued by the Australian National Telescope Facility \citep{Manchester} only 
a few are also $\gamma$-ray emitters. This could be due to strong selection effects as suggested by the 
existence of pulsars, such as Geminga, which are $\gamma$-ray loud pulsars, yet not observed in the radio. However, 
both the SNR and 
pulsar distribution give similar conclusions, as would any class of sources that follows a similar spatial distribution.

To draw more certain conclusions about the high energy $\gamma$-ray Galactic sky, new observations 
are of fundamental importance. VERITAS will be able to survey the same region of the sky as Milagro.
Milagro's recently improved sensitivity will better image the diffuse emission. Finally, 
GLAST will investigate the window of energy between 10 MeV to 300 GeV, covering the energy 
gap left between EGRET and the ground based gamma-ray observatories.

\begin{acknowledgments}
We would like to thank Andrew Strong, Igor Moskalenko and Nino Panagia for useful discussions and comments.
\end{acknowledgments}

\end{document}